# QUANTIZATION IN SPACETIME FROM NULL PATHS IN HIGHER DIMENSIONS

Paul S. Wesson[1,2]

[1]Dept. Physics and Astronomy, University of Waterloo, Waterloo, Ontario N2L 3G1, Canada

[2]Space-Time-Matter Consortium, http://astro.uwaterloo.ca/~wesson

Correspondence: Mail to (1) above. Email= psw.papers@yahoo.ca

Abstract

Massive particles on timelike paths in spacetime can be viewed as moving on null paths in a higher-dimensional manifold. This and other consequences follow from the use of Campbell's theorem to embed 4D general relativity in non-compactified 5D Kaluza-Klein theory. We now show that it is possible in principle to obtain the standard rule for quantization in 4D from the canonical metric with null paths in 5D. Particle mass can be wavelike, as suggested originally by Dirac, and other 4D/5D consequences are outlined.

1. Introduction

In modern versions of general relativity which employ an extra non-compactified dimension, particles stay on or close to 4D spacetime. In membrane theory, particles stay on the spacetime hypersurface because it is singular, so helping to explain their small masses in comparison to the Planck mass. In space-time-matter theory, particles stay close to spacetime because their motions are controlled by the cosmological constant, the extra dimension serving to induce 4D matter in accordance with Campbell's theorem. A review of both approaches is available [1], and here we do not propose to discuss their relative merits except to remark that both may be regarded as the low-energy limit of a theory in even higher dimensions which may unify gravity with the interactions of particle physics. However, both theories involve an extra force which exists in spacetime if there is relative motion between the 4D and 5D frames [2, 3]; and in both theories massive particles on timelike paths in spacetime may be viewed as moving on null paths in 5D [4, 5]. It is this latter property on which we will focus here. We will show that this condition leads to some remarkable conclusions, especially when it is combined with a change in the nature of the extra dimension from spacelike to timelike. It should be noted that such "two-time" metrics lead to acceptable physics in 5D, 6D and higher dimensions [6, 7], provided they are properly interpreted. Our interpretation will draw on certain technical results in the literature, particularly in regard to the cosmological 'constant' [8]. We will use the canonical metric of space-time-matter theory, rather than the warp metric of membrane theory, because it leads to a greater simplification of the field equations and the equations of motion [9-11]. However, these sets of equations are now known to be equivalent between the two approaches so we expect our results to have some generality.

We will show that for null paths in a 5D metric, the usual quantization rule in 4D is recovered. Also, a change in signature from $(+----)$ to $(+---+)$ causes the mass of a particle to become wavelike. This behaviour is reminiscent of that found by Dirac, who neatly classified the energy and ordinary momenta of a particle by embedding 4D deSitter space in 5D [12]. The use of embeddings to go from classical to quantum behaviour can be extended to higher dimensions.

2. From Classical to Quantum Mechanics: Wavelike Mass

We let upper-case Latin (English) letters run 0, 123, 4 for time, space and the extra coordinate, which we label $x^4 = \ell$. (This avoids confusion with the labels for spacetime.) The 5D line element in terms of its metric tensor is then $dS^2 = g_{AB}dx^A dx^B$. We let lower-case Greek letters run 0, 123 so the embedded 4D line element is $ds^2 = g_{\alpha\beta}dx^\alpha dx^\beta$. This leads to 4-velocities $u^\alpha \equiv dx^\alpha / ds$, which are normalized as usual via $u^\alpha u_\alpha = 1$ or 0 for massive and massless particles. The crux of the argument is that we can have $dS^2 = 0$ with $ds^2 \geq 0$ [4, 5]. This implies that in some sense all particles are in casual contact and photon-like in 5D. The extra coordinate which enables this should not be viewed as a regular coordinate in either the space or time manner: in membrane theory it determines the singular hypersurface of spacetime, while in space-time-matter theory it induces curvature and so matter in spacetime. In both cases, we are dealing with a local embedding of 4D in 5D. This procedure is governed by Campbell's theorem, which ensures that the 5D field equations based on the Ricci tensor $R_{AB}$ smoothly contain the 4D Einstein equations. These in terms of the Einstein tensor, the energy-momentum tensor and the cosmological constant are $G_{\alpha\beta} + \Lambda g_{\alpha\beta} = 8\pi T_{\alpha\beta}$ ($\alpha, \beta = 0,123$; we here absorb the gravitational constant $G$ and the speed of light $c$, and we will do the same with Planck's constant $h$, unless we need to make these explicit for physical clarity). Many solutions of the 5D field equations are known, including ones for the 1-body problem and cosmology which ensure agreement with observations [1]. However, we will not be concerned much with the field equations, but rather with the equations of motion. To put these into a useful form, we need to choose a system of coordinates or gauge.

The canonical gauge is the 5D analog of the 4D synchronous one of standard cosmology, and has a line element given by

$$dS^2 = (\ell/L)^2 g_{\alpha\beta}(x^\gamma, \ell) dx^\alpha dx^\beta - d\ell^2 \qquad . \qquad (1)$$

Here $L$ is a constant length, which by reduction of the field equations from 5D to 4D can be identified when $g_{\alpha\beta} = g_{\alpha\beta}(x^\gamma$ only) in terms of the cosmological constant. The precise relation is $\Lambda = 3/L^2$. The metric (1) is well-suited to ordinary mechanics, because it uses the 5 available degrees of coordinate freedom to remove the 4 potentials of electromagnetic type ($g_{4\alpha} = 0$) and flatten the potential of scalar type ($g_{44} = -1$). However, (1) is still general provided we allow that effects of the fifth dimension may be contained in the 4D sector via $g_{\alpha\beta} = g_{\alpha\beta}(x^\gamma, \ell)$. Indeed, it is this which leads to significant forms for the induced energy-momentum tensor $T_{\alpha\beta}$ in the space-time-matter approach. For massive particles it is known that when the metric has the
form (1) then $x^4 = \ell$ not only induces matter in 4D, but is actually a measure of particle rest
mass $m$. In a way, this is trivially clear from (1), whose 4D part contains the action element of
conventional mechanics *mds* if $\ell$ is proportional to $m$. However, we also come to the
interpretation $\ell = m$ in a more formal way, by evaluating the constant of the motion under
appropriate conditions for the time axis of (1). Of course, such an interpretation is lost for other
forms of the metric. This is an elementary example of how, in general, quantities will change
form under coordinate transformations which involve $x^4 = \ell$. It should be recalled, in this regard,
that the 5D transformations $\bar{x}^A = \bar{x}^A(x^B)$ and the 4D transformations $\bar{x}^\alpha = \bar{x}^\alpha(x^\beta)$ are not in
general equivalent. For example, the former preserves $R_{AB} = 0$ while the latter preserves $G_{\alpha\beta} = 0$.
We will note below how $\Lambda$ changes under a change of $\ell$; but for now we wish to focus on the
equations of motion when the metric has the canonical form (1). These may be obtained either
by extremizing the interval through $\delta[\int dS] = 0$ or by using the Lagrange approach. It is conven-
ient to give the laws of motion separately for spacetime and the extra dimension. We find:

$$\frac{d^2 x^\mu}{ds^2} + \Gamma^\mu_{\alpha\beta} \frac{dx^\alpha}{ds} \frac{dx^\beta}{ds} = f^\mu \qquad (2.1)$$

$$f^\mu \equiv \left(-g^{\mu\alpha} + \frac{1}{2} \frac{dx^\mu}{ds} \frac{dx^\alpha}{ds}\right) \frac{d\ell}{ds} \frac{dx^\beta}{ds} \frac{\partial g_{\alpha\beta}}{\partial \ell} \qquad (2.2)$$

$$\frac{d^2\ell}{ds^2} - \frac{2}{\ell}\left(\frac{d\ell}{ds}\right)^2 + \frac{\ell}{L^2} = -\frac{1}{2}\left[\frac{\ell^2}{L^2} - \left(\frac{d\ell}{ds}\right)^2\right] \frac{dx^\alpha}{ds} \frac{dx^\beta}{ds} \frac{\partial g_{\alpha\beta}}{\partial \ell} \qquad (2.3)$$

These relations show that the motion in spacetime is the standard geodesic one (where the
Christoffel symbol accounts for the 4D curvature) but modified by a fifth force $f^\mu$, which is
really an acceleration per unit (rest) mass. This force exists in both current approaches to 5D

relativity [2, 3]. It is zero by (2.2) if $d\ell/ds$ and/or $\partial g_{\alpha\beta}/\partial \ell$ is zero, showing that it arises from the relative motion between the 4D and 5D frames. It is therefore inertial in the Einstein sense. The general approach to (2) is by inserting a solution from the field equations into (2.3), using this to evaluate (2.2), and employing this to obtain the motion in spacetime from (2.1). Such an approach has been followed in several instances in the literature, and the effect of the extra term in (2.1) is tiny except in extreme situations such as the big bang. But while this method is necessary for $dS^2 \neq 0$, it is cumbersome and can be circumvented for $dS^2 = 0$ by using the metric (1) directly. Then for a 5D null-path we find that $\ell = \ell_* \exp[\pm(s-s_0)/L]$ where $\ell_*$ is a constant and $s_0$ is a fiducial value of the 4D proper time which we absorb. Returning to (2.3), however, we find that $\ell = \ell_* \exp(\pm s/L)$ renders the l.h.s. of that relation zero, and also the r.h.s. *irrespective* of the value of $\partial g_{\alpha\beta}/\partial \ell$. This means that such motion is generic, in the sense that it holds irrespective of the matter in spacetime. Also generically, it may be verified that the equations of motion (2) imply $d(\ell u^\mu)/ds = 0$ in the local limit. That is, the energy and 3D momenta of a particle are conserved with the identification $\ell = m$. We therefore proceed on the assumption that when the 5D metric has the canonical form (1), it represents a momentum manifold rather than a straight coordinate manifold, with the extra coordinate playing the role of particle rest mass.

The so-called shifted canonical gauge reveals further interesting properties of a 5D world. The canonical metric (1) was set up so that $\ell$ is locally orthogonal to the spacetime surface *s*, and has a last part $d\ell^2$ which leaves no ambiguity about the measurement of $x^4 = \ell$. (The analog of the proper distance in the fifth dimension is the same as the coordinate distance with a suitable choice of origin.) However, that last part in the metric is left unaffected by the shift $\ell \to (\ell - \ell_0)$, where $\ell_0$ is a constant. This may at first appear to be a trivial change. But recalling our comment above about changes in quantities under a change of coordinates that involves $\ell$, we again decompose the field equations, and find the remarkable result that the form of the cosmological constant is altered [8]. The shifted canonical metric has line element

$$dS^2 = \left[\frac{\ell - \ell_0}{L}\right]^2 g_{\alpha\beta}(x^\gamma, \ell) dx^\alpha dx^\beta - d\ell^2 \qquad . \qquad (3)$$

When this has $g_{\alpha\beta} = g_{\alpha\beta}(x^\gamma$ only), the 5D field equations give the effective 4D cosmological 'constant' as

$$\Lambda = \frac{3}{L^2}\left[\frac{\ell}{\ell - \ell_0}\right]^2 \qquad . \qquad (4)$$

This is in general variable, because $\ell = \ell(s)$. Only for $\ell \to \infty$ does it revert to its previous constant value $3/L^2$, and for $\ell \to \ell_0$ it can actually diverge. As regards $\ell(s)$, we could obtain this by using $g_{\alpha\beta} \to (1 - \ell_0/\ell)^2 g_{\alpha\beta}$ in (2.3), which is the appropriate conversion of (1) to (3). But as before, a more direct method is to put $dS^2 = 0$ in (3). A 5D null-path then corresponds to an $\ell$-path given by

$$\ell = \ell_0 + \ell_* \exp(\pm s/L) \qquad . \qquad (5)$$

Here as above, $\ell_*$ is a constant and the 4D proper time $s$ is defined via $ds^2 \equiv g_{\alpha\beta} dx^\alpha dx^\beta$, where we have again absorbed a fiducial value of it. We comment that while (5) is the general solution for $dS^2 = 0$, another solution of the type noted above may be obtained by solving (2.3) so that its l.h.s. and r.h.s. vanish simultaneously, irrespective of the value of $\partial g_{\alpha\beta}/\partial \ell$. This gives $\ell = \ell_0 \exp(\pm s/L)$, so by (4) $\Lambda = (3/L^2)[1 - \exp(\mp s/L)]^{-2}$, and the cosmological 'constant' decays from an unbounded value at the big bang ($s = 0$). However, in what follows we will use (5), because we are primarily concerned with particle physics. The thing we learn from the metric (3) is that the shift in the $\ell$-coordinate has the effect of resetting the particle mass as defined by the 4D action, and that the vacuum around it is also modified according to (4). Indeed, the latter shows that there is an intimate relation between the mass of a particle and the energy density of its attendant vacuum. This relieves the cosmological-'constant' problem, because $\Lambda$ as a measure of vacuum energy becomes a local parameter.

Another gauge sometimes considered in the 5D literature involves the change $\ell \to L^2/\ell$ in the canonical metric (1) and its analog for the shifted-canonical metric (3). To learn something from this inverted gauge, we need to consider just what type of mass we are discussing. The Equivalence Principle, properly expressed, states that gravitational mass (i.e. that involved in the force of gravity) is *proportional* to inertial mass (i.e. that involved in the energy inherent in an object). That is, $m_g$ is proportional to $m_i$ if we use the appropriate subscripts to denote the two types of mass. Now we can if we wish measure these two masses in a geometrical way by using the Schwarzschild radius $\ell_g = Gm_g/c^2$ and the Compton wavelength $\ell_i = h/m_i c$. Then the Equivalence Principle states that $\ell_g \ell_i =$ constant. But this is just the coordinate transformation of 5D relativity which we noted above, in the form $\ell_g \ell_i = L^2$. The point of this discussion is that it is essential to distinguish between gravitational and inertial mass, and their corresponding gauges. (For obvious reasons, the latter are sometimes referred to in 5D relativity as the Einstein and Planck gauges.) To appreciate why the distinction is important let us consider what happens if we assume that they are instead the same, with $Gm/c^2 = \ell = h/mc$. Then we obtain $m = (hc/G)^{1/2}$, the Planck mass of order $10^{-5}$ g. Real particles have masses many orders of magnitude less than this, a contradiction which has come to be known as the hierarchy problem. We learn from the foregoing that the transition from classical (i.e. gravitational) to quantum (i.e. particle) physics will probably involve a gauge that is related to the original canonical one by an inversion.

We therefore consider the inverted form ($\ell \to L^2/\ell$) of the canonical metric (1), when the 5D interval is null. Then we have

$$dS^2 = 0 = (L/\ell)^2 ds^2 - (L/\ell)^4 d\ell^2 \qquad . \qquad (6)$$

This gives the 'velocity' in the fifth dimension as $d\ell/ds = \pm \ell/L$, which is actually the same relation as follows from the straight-canonical metric (1). (The sign choice merely reflects the reversibility of the motion, and will henceforth be suppressed.) However, by the comments of the preceding paragraph, the physical measure for the mass is now via the Compton wavelength, or $\ell = 1/m$ (with the constants absorbed). Then we obtain

$$\int m ds = \int ds/\ell = L/\ell \qquad . \qquad (7)$$

The clear implication of this is that the action of particle physics is quantized as observed in 4D if there is structure in the fifth dimension such that $L/\ell = n$, an integer.

This is enlightening, but we would like to know the physical nature of the structure involved. Accordingly, we consider a new gauge, in which the extra dimension is timelike rather than spacelike. (This can be achieved by $\ell \to i\ell$ with $L \to iL$ in the above, and is effectively a Wick rotation of the extra coordinate.) The change in the signature of the metric from $(+----)$ to $(+---+)$ is analogous to that used in the Euclidean approach to 4D quantum gravity, and has been used previously in 5D relativity [6]. It can be carried out for any of the metric forms considered above, with similar results. For generality, we include a shift in $x^4 = \ell$ by $\ell_0$ and take the line element as

$$dS^2 = \left[\frac{L}{\ell - \ell_0}\right]^2 ds^2 + \left[\frac{L}{\ell - \ell_0}\right]^4 d\ell^2 \qquad . \qquad (8)$$

Then the 5D null-path $dS^2 = 0$ gives the orbit in the $\ell/s$ plane as

$$\ell = \ell_0 + \ell_* e^{\pm is/L} \qquad (9)$$

This describes a wave in $x^4 = \ell$ which oscillates around $\ell_0$ with amplitude $\ell_*$ and wavelength $L$. It is the complex analog of (5), and it is the complex nature of (9) which allows the 5D interval (8) to be null while allowing the 4D interval to remain finite.

Causality in spacetime is conventionally defined by the condition $ds^2 \geq 0$ on the 4D proper time $s$, and it is important to realize that this condition is preserved by the wave (9) even though the 5D interval is null with $dS^2 = 0$. That is, the first part of the 5D interval (8) is positive with real $s$ and $ds^2 > 0$ (for a massive particle), while the second part is negative due to the complex nature of (9), so allowing the 5D interval to be zero. Conventional causality is maintained for the 4D part of the metric, if the extra part is a complex wave. (We acknowledge that a complex coordinate is unconventional and discuss it in note 13.) By previous considerations we infer that the wave is 'supported' by the vacuum, whose energy density is fixed by the cosmological 'constant' $\Lambda$, in the usual interpretation where the vacuum fluid is taken as part of the 4D energy-momentum tensor with an effective equation of state $p_v = -\rho_v = -\Lambda/8\pi$ [1]. It should be noted in this regard that for metrics of canonical type, spacelike $x^4$ implies $\Lambda > 0$ while timelike $x^4$ implies $\Lambda < 0$ [8]. Also, the condition of a 5D null-path only fixes the behaviour of the fifth coordinate $\ell$ as a function of 4D proper time $s$. The appearance of $s$ explicitly in this manner is typical of dynamical problems in 5D, where the motion splits naturally into parts for spacetime and the fifth dimension [1]. The motion in 4D depends on the physics of the spacetime sector, and would be given in a specific case by a solution of the 4D part of the 5D geodesic equation, as in (2). In fact, a specific comparison of (9) with observation requires information about the nature of the spacetime hypersurface, including its topology. There are clearly many possible applications of (9), depending on the 4D physics, and these will need detailed investigation. However, all must share the property evident in (9), namely that there is a mass-related wave running across spacetime.

This conclusion is similar to the one arrived at in 1935 by Dirac [12]. As noted in Section 1, he obtained a natural identification for the energy and 3-momenta of a particle by embedding 4D deSitter space in a 5D spherical space, which implied that in general the mass should be a complex quantity. This agrees with (9), which says in essence that there is a gauge where the mass is wavelike. The most natural identification for (9), based on the action defined by the first part of (8), is that $m \equiv 1/(\ell - \ell_0)$. This implies that the observed mass is the r.m.s. one, given by

$$\overline{m} = \langle m^2 \rangle^{1/2} = 1/2\ell_* \qquad . \qquad (10)$$

This is the mass which would be measured by an observer using radiation whose wavelength is large compared to that of the mass wave (*L*), or a time period long compared to the equivalent period. Similar results follow from other forms of the canonical metric as discussed above. As above in (7), it can be shown that the action $\int mds$ is quantized in terms of an integer *n* that is a ratio of lengths. These depend on whether the 4D spacetime hypersurface is open or closed. (In the flat case, $n = L/(\ell - \ell_0)$; but since $\Lambda < 0$ for a pure-canonical metric with a timelike extra dimension, it is also possible to consider an $\ell$-wave that runs around a compact spacetime with a phase angle $d\theta = ds/\ell$, which leads to a value for *n* that essentially counts the number of revolutions.) We will present results on these different spacetime options in a future work. Irrespective of the details, the realization that particle rest mass can be viewed as a wave like (9) resolves some long-standing problems to do with wave-particle duality. For example, the double-slit experiment has come to be regarded as a quaint curiosity, but it still represents a hole in our understanding of matter. This hole may be plugged by the realization that particle and wave behaviour can be viewed as isometries, in the sense that they correspond to a 5D manifold with spacelike and timelike extra dimensions, respectively. It is obvious that if particle rest mass *m* can be wavelike, then the energy ($mu^o$) and 3-momenta ($mu^{123}$) also become waves, as first expounded by deBroglie. It is not known if waves of the type we have discussed can show interference (see the preceding paragraph). Clearly, more work is required from the theoretical side. Insofar as we have neglected the electromagnetic potentials ($g_{\alpha 4}$) in our metrics, neutron interferometry is the indicated way to investigate our concept of wavelike mass from the observational side.

3. Conclusion

In five-dimensional noncompactified general relativity, massive particles may travel on 5D null paths while on 4D timelike paths ($dS^2 = 0, ds^2 \geq 0$). Particles are in causal contact in 5D while they appear distinct in 4D spacetime. This opens up a new vista, where the 5D manifold can be used to explain several previously-puzzling aspects of 4D dynamics while doing no violence to the latter. Thus we can explain the usual quantization rule $\int mc ds = n$, and understand why an object can appear as both a particle and a wave.

Five-dimensional relativity is currently popular in the forms of membrane theory and space-time-matter (or induced-matter) theory. Both are in agreement with macroscopic observations, because Campbell's theorem guarantees a 5D embedding for 4D general relativity [1]. In the present work, we have taken several metrics commonly used in cosmology and applied them to particle physics. The canonical metric (1) has a length scale set by the cosmological constant, which may also be viewed as fixing the energy density of the local vacuum. Its associated equations of motion (2) involve a fifth force or acceleration, which is compatible with the conservation of momentum, when however the mass is generally variable. Indeed, that metric is set up so that the extra coordinate is a measure of the rest mass of a particle. The canonical metric can be transformed by altering the fifth coordinate to produce other useful gauges. The shifted form (3) results in an expression for the cosmological 'constant' (4) that depends on the extra coordinate. (The last equation predicts that the cosmological 'constant' scales for an interaction as the inverse-square of the energy, which might be testable using the Large Hadron Collider.) The extra coordinate or mass varies as per (5). The inverted canonical metric (6) leads directly to quantization with the standard rule (7). This is essentially because the 4D action is a dimensionless ratio of 5D lengths, provided that the mass measure involves Planck's constant. The structure in 5D needed to account for 4D quantization can be elucidated by using a form of the canonical metric (8) with a timelike extra dimension. Wavelike behaviour ensues for the fifth coordinate, specified by (9). This results in the mass being also wavelike; though even if the oscillation is around zero, the r.m.s. or measured value of the mass is finite and related to the amplitude, as in (10). The wavelike nature of particle mass is in formal agreement with an old result of Dirac [12]. If mass is wavelike, so are the energy and momenta. We then readily understand the origin of deBroglie waves, and realize that geometric isometries may explain wave-particle duality. The most appropriate technique for experimental investigation of the ideas set forward here would appear to be neutron interferometry.

Our account has been exploratory, and many issues remain to be clarified by detailed work. Encouragingly, the 5D approach avoids both of the problems which currently bedevil the unification of gravity and particle physics. The cosmological-'constant' problem is sidestepped, because that parameter is now a local measure of variable vacuum energy, not a universal constant. The hierarchy problem does not arise, because gravitational and inertial mass are now distinct, with gauges for their measurement that are not mixed, so the Planck mass is not realized in nature.

In conclusion, we have in a provisional way constructed a framework which, at the expense of adding an extra dimension, allows us to see classical and quantum mechanics as two aspects of the same geometry.

Acknowledgements

This work grew out of earlier collaborations with other members of the Space-Time-Matter consortium (http://astro.uwaterloo.ca/~wesson).  It was supported in part by N.S.E.R.C.

References

[1]  P.S. Wesson, Five-Dimensional Physics (Singapore: World Scientific, 2006).

[2]  D. Youm, Phys. Rev. D 62, 084002 (2000).

[3]  P.S. Wesson, B. Mashhoon, H. Liu and W.N. Sajko, Phys. Lett. B 456, 34 (1999).

[4]  D. Youm, Mod. Phys. Lett. A 16, 2371 (2001).

[5]  S. S. Seahra and P.S. Wesson, Gen. Rel. Grav. 33, 1731 (2001).

[6]  P.S. Wesson, Phys. Lett. B 538, 159 (2002).

[7]  I. Bars and Y.-C. Kuo, Phys. Rev. Lett. 99, 041801 (2007).

[8]  B. Mashhoon and P.S. Wesson, Class. Quant. Grav. 21, 3611 (2004); Gen. Rel. Grav. 39, 1403 (2007).

[9]  J. Ponce de Leon, Mod. Phys. Lett. A 16, 2291 (2001).

[10]  J. Ponce de Leon, Int. J. Mod. Phys. D 11, 1355 (2002).

[11]  P.S. Wesson, Class. Quant. Grav. 19, 2825 (2002).

[12]  P. A. M. Dirac, Ann. Math. 36, 657 (1935).

[13] A fifth coordinate which measures particle rest mass and has the nature of a complex wave is unconventional from the standpoint of 4D general relativity. In the latter, the metric is usually taken with signature ($+---$) and the coordinates for time ($x^0$) and space ($x^{123}$) are taken to be real. On occaision, $x^0$ is changed to an imaginary quantity in order to yield a Euclidean metric which helps with quantization (see G.W. Gibbons, S.W. Hawking, Phys. Rev. D15, 2738, 1977 and S. Coleman, Phys. Rev. D15, 2929, 1977). This is acceptable mathematically, but complicates the nature of causality physically, a problem which affects other approaches to quantization with complex quantities (see J. Plebanski, J. Math. Phys. 18, 2511, 1977 and A. Ashtekar, Phys. Rev. Lett. 57, 2244, 1986). As regards the metric, it is somewhat artificial to distinguish between a complex coordinate and a complex metric coefficient, and the latter appears in otherwise well-behaved wave-like solutions of the 5D field equations (see A. Billyard, P.S. Wesson, Gen. Rel. Grav. 28, 129, 1996). Also, a complex coordinate in 5D can always be rewritten in terms of real coordinates in 6D (see I. Bars, Y.-C. Kuo, Phys. Rev. Lett. 99, 041801, 2007 and I. Bars, C. Deliduman, D. Minic, Phys. Rev. D59, 125004, 1999). In the present approach, we take the formal view that equation (9) is a solution of equation (8) with null 5D interval $dS^2$, and does not alter the nature of 4D causality as defined by $ds^2$. This may be verified by a simple substitution of (9) into (8). It should be noted that we have refrained from writing the exponential term in (9) as the sum $\ell_*[\cos(s/L)+i\sin(s/L)]$ and taking the real part, because strictly speaking this is only valid for a linear form, and there are questions about the physical interpretation of (9) as mentioned in the main text. However, irrespective of the details of how (9) is interpreted, it is remarkable that it parallels results of Dirac (Ann. Math. 36, 657, 1935) and reproduces other properties of particle mass not usually recoverable from a classical approach.